\documentclass{aa}
\usepackage{indentfirst}
\usepackage{braket}
\usepackage{graphicx}
\usepackage{txfonts,natbib}
\usepackage{amsmath,amssymb,mathtools}
\usepackage{subcaption}
\usepackage{soul}
\usepackage{threeparttable}
\usepackage[colorlinks,citecolor=blue]{hyperref}

\makeatletter
\def\@biblabel#1{}
\makeatother

\begin{document}

   \title{Is the primary CoRoT target HD43587 \\ under a Maunder minimum phase?}

   \subtitle{}

   \author{R.R. Ferreira \inst{1},
           R. Barbosa \inst{2},
           M. Castro \inst{1}, 
           G. Guerrero \inst{2}, 
           L. de Almeida \inst{1}, \\
           P.  Boumier \inst{3}
           \and
           J.-D. do Nascimento Jr. \inst{1,4}
           }

   \institute{Departamento de Física Teórica e Experimental, Universidade Federal do Rio Grande do Norte, Natal, RN, CEP: 59072-970, Brazil\\
              \email{rafaelferreira@fisica.ufrn.br}
              \and 
              Departamento de Física, Universidade Federal de Minas Gerais, Av. Antonio Carlos, 6627, Belo Horizonte, MG, CEP: 31270-901, Brazil
            \and
             Institut d’Astrophysique Spatiale, UMR8617, Université Paris-Saclay, CNRS, 91405 Orsay, France
             \and
             Harvard-Smithsonian Center for Astrophysics, Cambridge, MA 02138, USA
             }

   \date{Received 29 November 2019 / Accepted 1 June 2020}

 
  \abstract
   {One of the most enigmatic phenomena related to solar activity is the so-called Maunder minimum phase. It consists in the lowest sunspot count ever registered for the Sun and has not been confirmed for other stars to date. Since the spectroscopic observations of stellar activity at the Mount Wilson Observatory, the solar analog HD43587 has shown a very low and apparently invariant activity level, which makes it a Maunder minimum candidate.}
   {We aim to analyze the chromospheric activity evolution of HD43587 and its evolutive status, with the intention of unraveling the reasons for this low and flat activity.}
   {We used an activity measurements dataset available in the literature, and computed the S-index from HARPS and NARVAL spectra to infer a cycle period. Additionally, we analyzed the CoRoT light curve of HD43587, and applied gyrochronology and activity calibrations to determine its rotation period. Finally, based on an evolutionary model and the inferred rotation period, we used the EULAG-MHD code to perform global MHD simulations of HD43587 to get some insight into its dynamo process.}
   {We confirm the almost flat activity profile, with a cycle period $P_{\mathrm{cyc}} = 10.44\pm3.03$ yrs deduced from the S-index time series, and a long-term trend that might be a period of more than 50 yrs. It was impossible to define a rotation period from the light curve, however gyrochronology and activity calibrations allow us to infer an indirect estimate of $\overline{P}_{\mathrm{rot}} = 22.6 \pm 1.9$ d. Furthermore, the MHD simulations confirm an oscillatory dynamo with a cycle period in good agreement with the observations and a low level of surface magnetic activity.}
   %
   {We conclude that this object might be experiencing a "natural" decrease in magnetic activity as a consequence of its age. Nevertheless, the possibility that HD43587 is in a Maunder minimum phase cannot be ruled out.}

\titlerunning{Is the primary CoRoT target HD43587 under a Maunder minimum phase?}
\authorrunning{R.R. Ferreira et al.}
\maketitle
\section{Introduction}

The study of the sunspot number as well as that of solar variability have been        
good indicators of magnetic activity \citep{eddy,stuiver,kivelson}. Continuous
records of the sunspot number have existed since the first observation by Galileo at
the beginning of the seventeenth century\footnote{Available at the Royal Observatory of
Belgium:  {\url{http://www.sidc.be/silso/datafiles}}}. The solar activity from
observational data exhibits a period of about 11 years with amplitude variations
that remain an open issue in modern astrophysics \citep{weissthomas}.

One of the most intriguing phenomena observed in the sunspot counting survey is the long
period phase, posteriorly called the Maunder Minimum (MM hereafter) when almost no sunspots were observed on the solar surface \citep{eddy}. It lasted for about 70 yrs between the sixteenth and seventeenth centuries. There is still no complete explanation for this period of extremely low activity \citep{wright,lubin}, however, it is historically known that it coincided with a short glacial era on Earth \citep{eddy}.

The fundamental question about how singular the magnetic activity evolution of the Sun
has motivated long-term surveys of stellar activity of solar-type stars. Chromospheric stellar activity basically consists of the temporal variations of the
stellar magnetic field caused by mass loss and radiation outside the star \citep{hall}. These variations have a connection with the flux of Ca II H\&K lines, which have been widely used as a spectroscopical indicator of stellar activity \citep{auer,skumanich,wilson1968}. Regarding the Sun, for instance, \cite{wilson63,wilson64,wilson1968} established the connection between solar activity and the strength of the Ca II H\&K lines.

Strong efforts have been made over the last decades seeking to increase the
spectroscopic data \citep{schwarzschild,auer}. In his seminal work, \cite{skumanich72}
showed that activity, as given by the Ca II H\&K lines, rotation, and lithium abundance, decay as the inverse square root of time. This paper was a milestone in stellar chromospheric activity studies. The first observational program to continuously measure the activity of main sequence stars, the HK project at the Mount Wilson Observatory\footnote{Located at Mount Wilson, at 1742m altitude in the San Gabriel Mountains, LA, California, the observatory has two telescopes; the Hale telescope (1.5 m), built in 1908, and the Hooker telescope (2.5 m), built in 1917.} (MWO hereafter), was described in detail by \cite{vaughan1978}. Improved bases of the project and its results of spectroscopical measurements were related by \cite{wilson1978}. Later on, \cite{duncan} presented a large set of time series spectroscopical measurements from the Mount Wilson survey containing data for 1296 stars observed between 1966 and 1983. \cite{baliunas95} presented enhanced and extended results of Mount Wilson stars and flagged the activity of the stars as cyclic and flat.

From these observational results, it is accepted that young stars have strong activity levels \citep{donascimento}. However, some young stars present extremely low activity over many years. This unexpected phase is a puzzle for theoreticians and observing astronomers. Many authors have proposed using the solar chromosphere as a pattern to study the activity evolution of other solar-type stars. It could be an approach to investigate the occurrence of MM in stars similar to the Sun
\citep{schrijvera,schrijver1989,schrijverc,schroder}.

In this work, we investigated the low spectroscopic activity of the solar analog HD43587 over 50 years of measurements. This object has been widely referred to as a flat activity star and an MM candidate \citep{baliunas95,wright,lubin,schroder}. We used
the fluxes in the Ca II H\&K lines, through the Mount Wilson S-index ($S_{MW}$), whose photometric measurements were taken with the 1.5m telescope at MWO. Since the MWO survey, this object has been observed by several others surveys (detailed below in the Section \ref{HD43587}), as well as being described by \citet{wright}, \citet{morel}, and \citet{radick2018}. We focused on the continuous low activity level presented from spectroscopic measurements, showed by the time-averaged value $\braket{S_{MW}}$ = 0.154. In addition, the observations from the CoRoT satellite provide a photometric activity proxy $S_{ph}$ = $102 ppm$ as described by \citet{mathur2014}, sustaining the flat activity status. 

This paper is organized as follows; in Section 2 we briefly present the observational data and reduction used along this work, and in Section 3 we present our results. Section 4 shows an attempt to simulate the star HD43587 by using the EULAG-MHD code. Finally, we present our conclusions in Section 5.

\section{Observational data}
\label{data}

In this section, we briefly present the characterization of the solar analog HD43587, and explain all of the spectroscopic data used, as well as the $S_{MW}$ calibration for each instrument. In the last subsection, we explain the photometric data from the CoRoT satellite.

\begin{figure}[h]
\centering
\includegraphics[width=9cm,height=9cm]{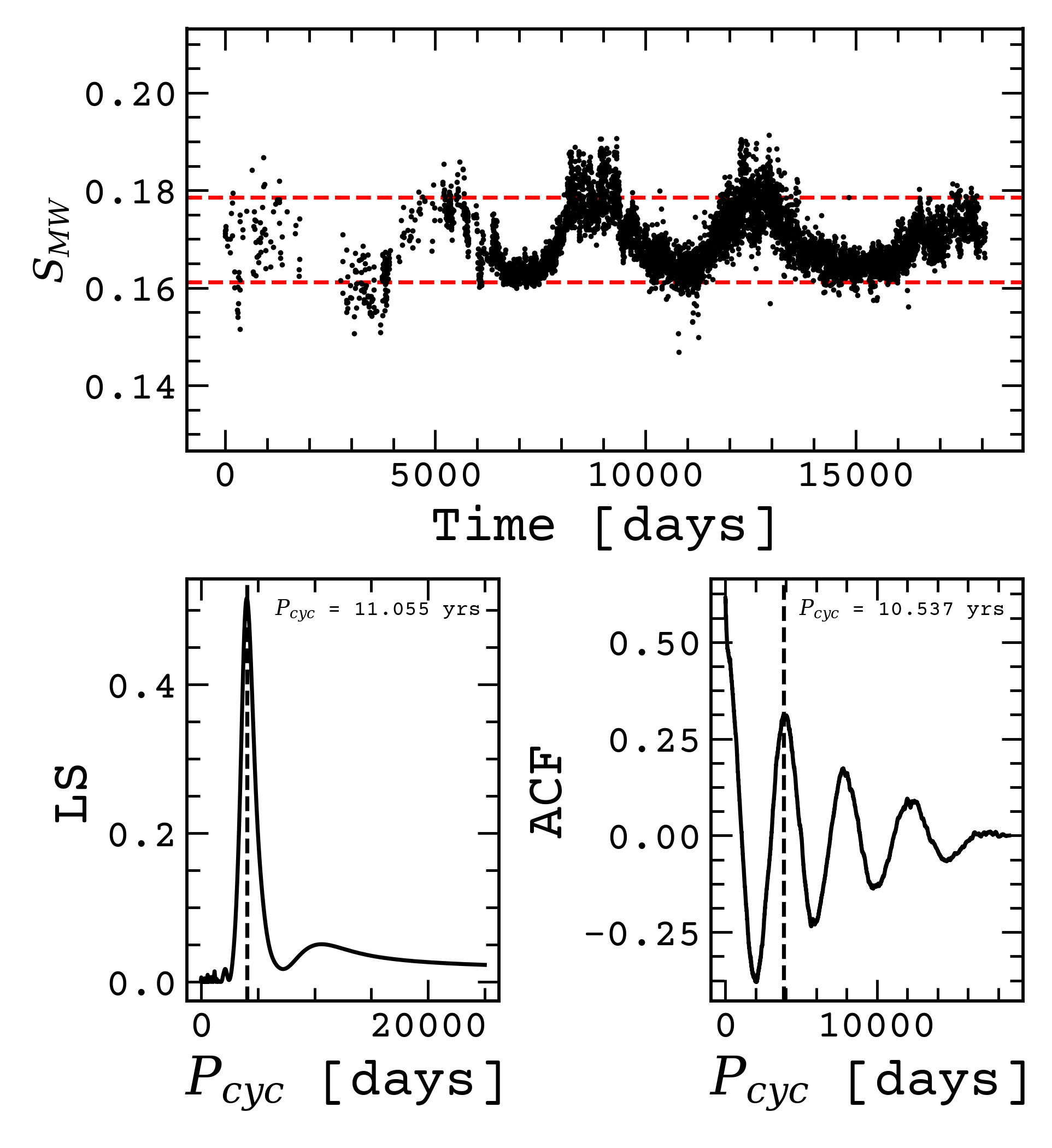}
\caption[Sun timeseries]{\emph {Sun's chromospheric activity over 50 years according to \cite{egeland}. Upper: complete time series measurements made over 50 years depicting the $S_{MW}$ periodic profile. Bottom: both the Lomb-Scargle (left), and the autocorrelation function \citep[ACF, see ][]{brockwell2002} (right) periodograms showing the $P_{\mathrm{cyc}}$$\approx$11 yr solar activity cycle.}}
\label{fig:sun}
\end{figure}

\subsection{The solar analog HD43587} 
\label{HD43587}

Our object HD43587 (more precisely HD43587Aa) belongs to a quadruple system composed
of two distant main-sequence visual binaries and localized in the Orion constellation. It is a bright sun-like star, and its fundamental parameters are presented in Table \ref{table:1}. This system has been largely observed by many authors \citep{duncan,baliunas95,catala,morel}. This star was also monitored by using the CFHT (Canadian-France Hawai Telescope), for four seasons: January 2002 (one night), June 2002 (two nights), September 2004 (one night) and September 2005. In all cases, $H_{2}$ and [Fe II] filters \citep{catala} were used.  

HD43587 was a CoRoT primary target in the seismology program with ID CoRoT 3474, observed during the LRa03 for 145 days. Complementary observations were done with the spectrograph High Accuracy Radial velocity Planet Search (HARPS), mounted on the 3.6m telescope in La Silla Observatory/ESO \citep{boumier}. In addition, it has been monitored by several spectroscopical long-term observational surveys dedicated to activity, such as the HK project of MWO \citep{duncan,baliunas95} between 1966 and 1990, and the California and Carnegie Planet Search, at the Keck Observatory between 1995 and 2000 \citep{wright}. This object was also continuously monitored by Automatic Photometric Telescopes of the Tennessee State University at the Fairborn Observatory, and the Solar-Stellar Spectrograph at the Lowell Observatory between 1995 and 2016 \citep{radick2018}. It was also  observed by the NARVAL spectrograph, mounted on the Bernard Lyot Telescope at the Observatoire du Pic du Midi between 2009 and 2010\footnote{\url{http://polarbase.irap.omp.eu}} \citep{petit}. Spectroscopic analysis of this star performed by the McDonald Observatory 2.1m Telescope and Sandiford Cassegrain Echelle Spectrograph derived a projected rotation velocity $v~sin~i = 4.9$ km.s$^{-1}$ \citep{Luck2017}.

\begin{table*}[h]
\caption{Fundamental parameters of HD43587 according to some authors}             
\label{table:1}      
\centering                          
\begin{tabular}{c c c c c}        
\hline\hline                 
 & Morel et al. (2013) & Boumier et al. (2014) & \multicolumn{2}{c}{Castro et al. (2020)} \\    
 & & & TGEC & CESTAM \\
\hline                        
   Mass [$M_{\odot}$] & 1.049 $\pm$ 0.016 & 1.04 $\pm$ 0.01 & $1.020\pm0.004$ & $1.04 \pm 0.01$  \\      
\hline   
   Radius [$R_{\odot}$] & 1.15 $\pm$ 0.01 & 1.19    & $1.19\pm0.01$ & 1.18  \\
\hline
   [Fe/H] [dex] & -0.02 $\pm$ 0.02 & 0.01     & $-0.026\pm0.003$ & 0.025  \\
\hline
   Age [Gyr] & 4.97 $\pm$ 0.52 & 5.60 $\pm$ 0.16   & $6.76\pm0.12$ & $5.7 \pm 0.1$ \\
\hline
   $T_{eff}$[K] & 5947 $\pm$ 17   & 5951 &  $5952\pm27$ & 5979 \\
\hline                                   
\end{tabular}
\end{table*}

This star has been considered by various authors as a strong MM candidate \citep{baliunas95,lubin,schroder,morel}. Its long-time activity series shows a long flat activity lasting about 50 years. Its time-averaged activity index is $\braket{S_{MW}} = 0.154 \pm 0.004$, while for the Sun, $\braket{S_{MW}}_{\odot} = 0.167 \pm 0.006$. 
The observational spectroscopic data considered here were obtained in part from already published long-term surveys, as well as from HARPS and NARVAL spectra. These data are detailed as follows:

\begin{itemize}
    \item NSO/MWO: measurements from MWO obtained between 1966 and 1983 \citep{duncan}. Additionally, we mainly used a dataset updated from National Solar Observatory (NSO) with S-index measurements until 1995\footnote{{\url  {https://www.nso.edu/data/historical-data/mount-wilson-observatory-hk-project/}}};
    
    \item NARVAL: 50 high-resolution spectra obtained in the archive available from Narval/Polarbase \citep{petit};
    
    \item HARPS: three high-resolution spectra from HARPS/ESO \citep{morel};
    
    \item  CCPS: S-index activity measurements from the California and Carnegie Planet Search (CCPS) program at Keck Observatory, available in \cite{wright}.
    
    \item LO: a dataset of chromospheric Ca II H+K emission observations between 1994 and 2016 from the Lowell Observatory (LO), available in \cite{radick2018}.
\end{itemize}

\begin{figure*}[h]
\centering
\includegraphics[width=\textwidth]{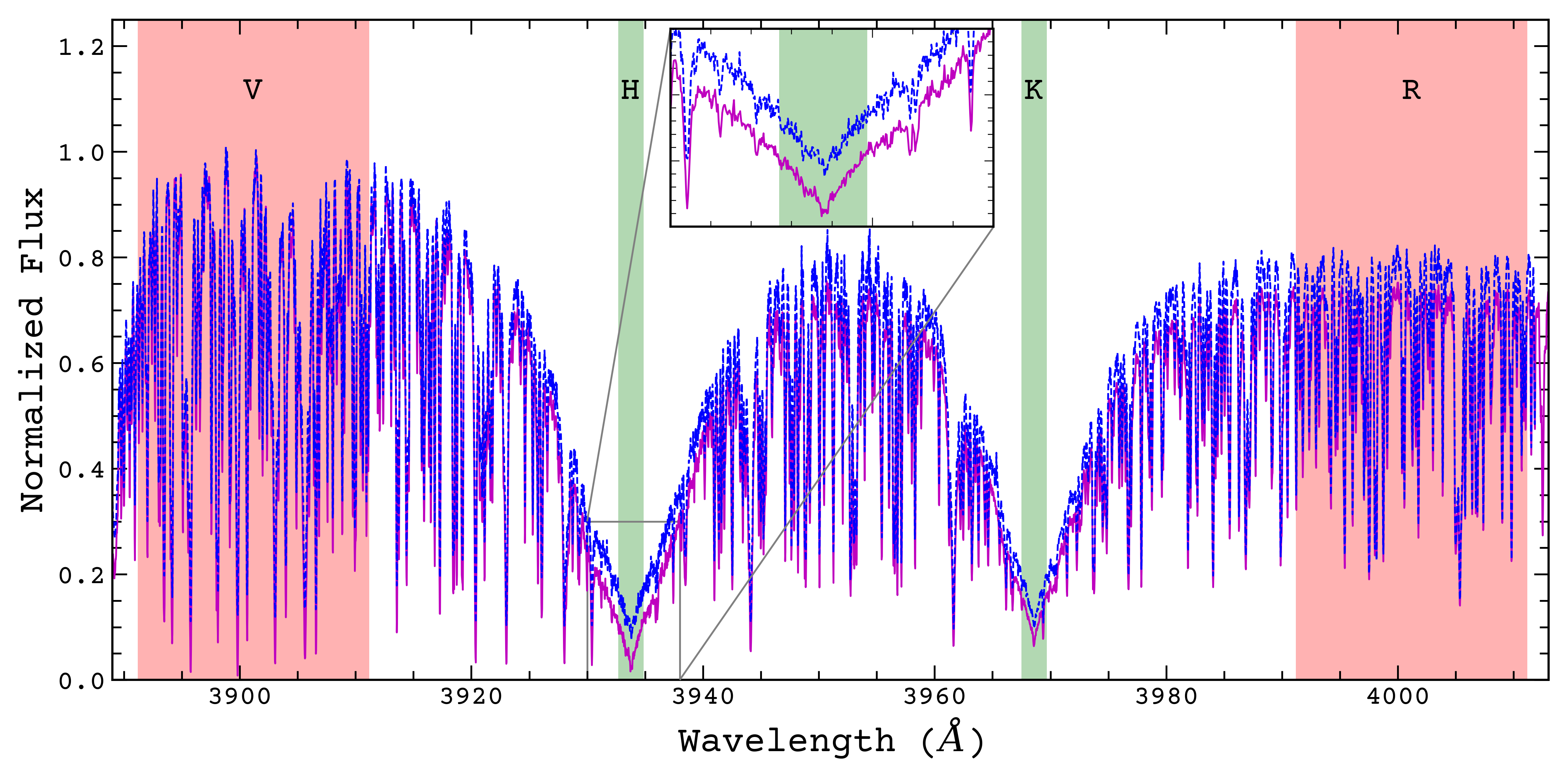}
\caption[HD43587spectra]{\emph {HD43587 spectra from NARVAL spectrograph. The magenta spectrum corresponds to the maximum $S_{MW}$ value of the dataset, observed on February 13, 2010. The blue spectrum corresponds to the minimum $S_{MW}$ value, observed on January 18, 2010. Both spectra were normalized from $3889$ {\AA} to $4013$ {\AA}. An upward shift in flux of 0.05 was applied in the blue spectra to better visualization. The inset panel shows the region between 3930 {\AA} and 3938 {\AA} in detail. }}
\label{fig:spectra}
\end{figure*}

Recently, \cite{castro} presented an evolution status analysis of this star using an optimization approach, by modelling through two different evolution codes, TGEC  \citep{huibonhoa} and CESTAM \citep{marques}, and by using spectroscopic, photometric, and seismic constraints. Results of the optimization are presented in Table \ref{table:1}. These results confirm the solar analog status of HD43587, and indicate a star that is slightly more massive and older than the Sun, which is in agreement with its observed low activity level.

\subsection{The Mount Wilson S-index}

Long-term observational programs at the MWO have been relevant to the detection of anomalous behaviour in stellar activity like the extreme low-activity cases, for instance, young solar-like stars probably in MM phase. In fact, the MM has been investigated by many authors since the 70s \citep{eddy,duncan,baliunas95,wright,lubin,schroder,radick2018}, initially in the MW project and afterward in other surveys. Over the years several indexes to measure stellar activity have been proposed. Here, we adopt the well-known MW S-index (see Eq.~\ref{eq1}), used to measure the stellar chromospheric activity and to investigate the stellar activity evolution.

Stellar spectra provide valuable information about stellar chromospheres. To measure stellar chromospheric activity, we used the flux in the Ca II H\&K lines, to derive the $S_{MW}$ index, which was initially proposed in the MW project \citep{duncan,baliunas90,baliunas95}.

The $S_{MW}$ is computed as follows:
\begin{eqnarray}
    \centering
    S_{MW} &=& \alpha \frac{H + K}{R + V},
    \label{eq1}
\end{eqnarray}

\noindent
where $\alpha$ is an observational constant defined by \cite{duncan}, $H$ and $K$ mean the flow strengths in triangular bandpasses with FWHM = 2.18 and centralized at 3933.633 {\AA} (K) and 3968.469 {\AA} (H), and $R$ and $V$ are continuous rectangular bands centralized at 3901.07 {\AA} (V) and 4001.07 {\AA} (R). \\

Since 1960, many authors have investigated the spectroscopic activity using S-index time series \citep{duncan,baliunas90,wright}. 
Few stars with low activity or flat activity, compared to the Sun at the minimum, have been reported. The value of the time average $\braket{S_{MW}}$ for these objects is slightly smaller than $\braket{S_{MW}}_{\odot}$. Subsequently, some studies were performed to understand the nature of flat-activity stars, among them, \cite{schroder} compared the solar chromospheric activity as a pattern in the context of solar-like stars. In spite of these efforts, it has been difficult to precisely determine any star in the MM phase. 

Since the first spectroscopic activity measurements, the solar S-index has varied between 0.16 (minimum) and 0.22 (maximum). The S-index time series of the Sun from \cite{egeland} is shown in Fig. \ref{fig:sun}. According to \cite{schroder}, the low activity levels of the Sun are a good pattern for investigate the activity evolution of other solar-like stars under a long flat-activity profile. These time-lapses of lower activity are useful to study the chromospheric flat-activity phase in solar-type objects like HD43587. This may be possible evidence of flat-activity stars at the MM phase, as proposed by \cite{baliunas95}.  

\subsection{S-index from NARVAL spectra}

We also used spectra obtained with the NARVAL spectrograph, mounted on the Telescope Bernard Lyot (TBL) located at Pic du Midi, France. The spectra are available in the PolarBase\footnote{\url{http://polarbase.irap.omp.eu/}} archive \citep{petit}. We used 50 high-resolution spectra collected in September 2009, January 2010, and February 2010. It consists of five spectra per night, totalizing 10 nights. The spectra corresponding to the maximum and minimum of the activity observed during this campaign are shown in Fig. \ref{fig:spectra}. 

Since different spectrometers have different bandpasses, it is necessary to use a different calibration to derive the $S_{MW}$-index for each instrument. For NARVAL spectra, $S_{MW}$ is derived following the method proposed by \cite{marsden}, where a non-renormalized spectrum was used, similar to that used by \cite{wright}. The activity index is computed as follows:

\begin{eqnarray}
\centering
S_{MW} &=& \frac{12.873H + 2.502K}{8.877R + 4.271V} + 1.183x10^{-3}.
\label{eq.2}
\end{eqnarray}

The coefficients in Equation (\ref{eq.2}) were obtained using a least-square fit \citep[see Fig. 4 in ][]{marsden} to the
113 stars observed by NARVAL \citep{marsden} and by the California Planet Searcher \citep{wright}.

\subsection{S-index from HARPS spectra}

We used three high-resolution spectra observed between December 2010 and January 2011, obtained with the spectrograph HARPS, in the high-efficiency mode \citep[for detailed information, see Section 2 of ][]{morel}. These spectra are also available in the SISMA database \citep{rainer}.

We computed the instrumental S-index from HARPS spectra following the same calibration as prescribed by \citet{lovis}:

\begin{eqnarray}
\centering
S_{HARPS} &=& \alpha \cdot 8 \cdot \frac{1.09{\AA}}{20{\AA}} \cdot \frac{H + K}{R + V}.
\label{eq.3}
\end{eqnarray}

\noindent
The calibration to the MW scale was done using a linear calibration between both indexes obtained with standard stars. According to \cite{lovis}, the activity index transformation is

\begin{eqnarray}
\centering
S_{MW} &=& 1.111 \cdot S_{HARPS} + 0.0153.
\label{eq.4}
\end{eqnarray}

\subsection{Photometric observation from CoRoT mission}

The CoRoT satellite observed the object HD43587 over 145 days \citep{boumier}, during the third long run (LRa03). The LRa03 
was in the direction of the Galactic anticenter and was performed
from October 2009 to March 2010. We made all reductions from the light curve extracted from the level 2 official products \citep{Samadi2007}. The possibility that the clear absence of a rotational modulation is due to the fact that the star is seen pole on, is ruled out by the determination of the projected rotation velocity by \citet{Luck2017} and mentioned in Sect.~\ref{HD43587}.

\begin{figure*}[h]
\centering
\includegraphics[width=\textwidth]{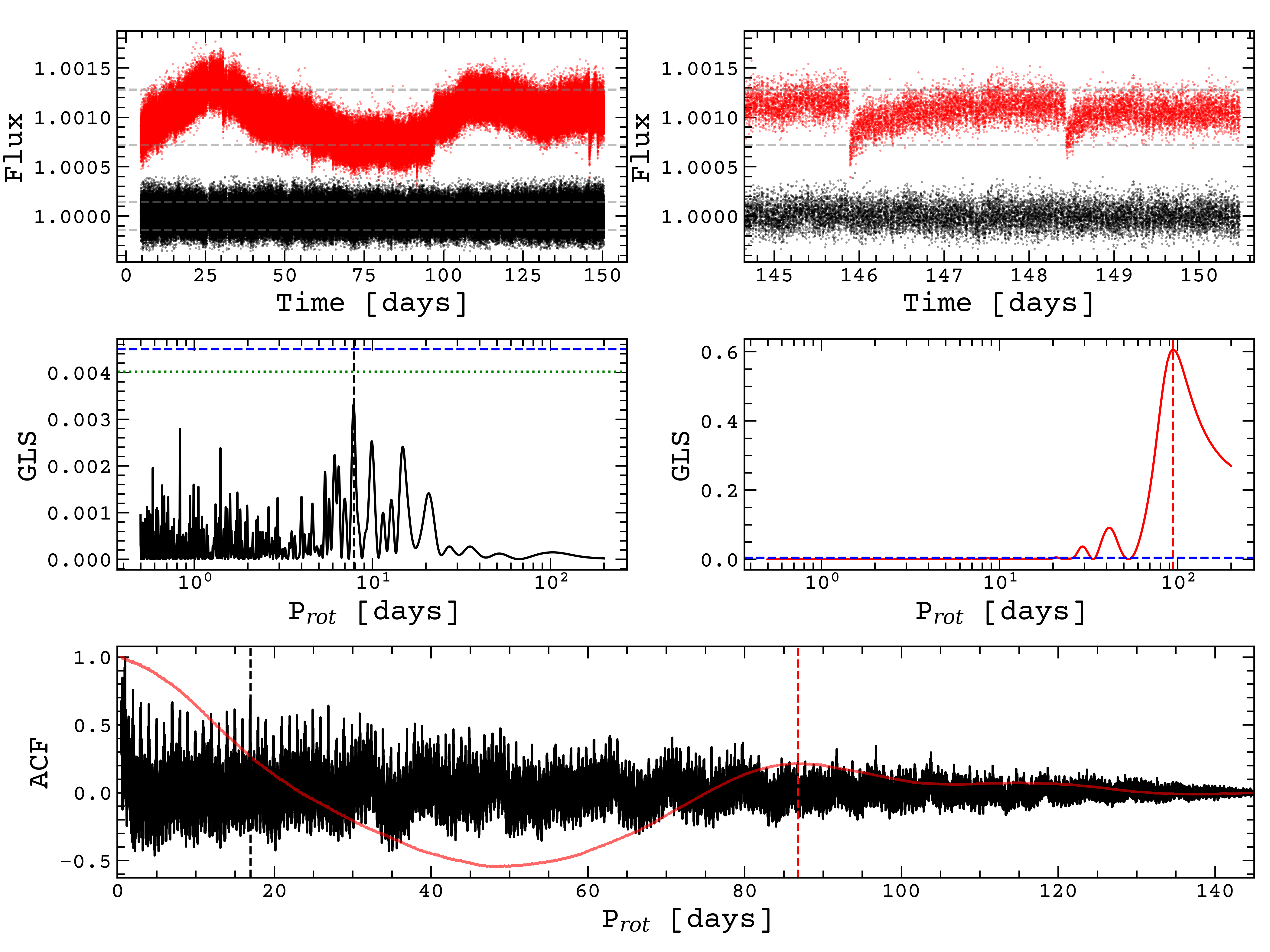}
\caption[HD43587LC]{\emph{CoRoT light curve of HD43587. Top-left panel: raw light curve including discontinuities (red dots) and the detrended light curve (black dots). Top-right panel: detail of the discontinuities at the end of the light curve (red dots) and the flat aspect of the curve after the reduction (black dots). Middle-left panel: GLS periodogram for the detrended light curve with false alarm probability of 5$\%$ and 1$\%$ (green dots and blue dashed line, respectively). Middle-right panel: GLS periodogram for raw light curve with false alarm probability the same as the detrended analysis. Bottom panel: ACF periodogram for the same two reductions.}}
\label{fig:HD43587LC}
\end{figure*}

\section{Analysis and discussion}

This section is divided in three parts. The first one presents the analysis of the photometric data collected by the CoRoT satellite. The second part presents the long-term time series analysis of the $S_{MW}$ and the inferred activity cycle. In the third part, we discuss about the indirect determination of the HD43587 rotation period.

\subsection{The HD43587`s CoRoT light curve}
\label{HD43587_light_curve}

Based on the photometric data from the CoRoT satellite, we investigated the variability of the light curve of HD43587. Aiming to determine the rotational period, we computed the Generalized Lomb-Scargle (GLS) periodogram \citep{gls} as well as the Autocorrelation Function \citep{brockwell2002} (see Fig.~\ref{fig:HD43587LC}). We used the "DATEJD" as the time array and the "RAWFLUX" in electrons per second as the raw flux to create the light curve. The error relative to the raw flux, "RAWFLUXDEV" in electrons per second, was used  to compute $S_{ph}$, as explained below. We also took into account the "RAWSTATUS" flag to make sure we used valid data. The resulting valid data exhibits a significant jump in the flux in the first 4.65 days. This discontinuity creates a false tendency on the overall aspect of the light curve, and, even when corrected, this discontinuity 
does not have the same dispersion in flux as the rest of the light curve. Thus, we decided to completely remove the first 4.65 days of the data. Then, we followed two approaches. First, we tried to recover the GLS and ACF periodograms (red lines in the lower panels of Figure \ref{fig:HD43587LC}) from the raw light curve (after removing the jump). We initially obtained the periodicities $P = 93.98^{101.31}_{-43.49}$~days with the GLS and $P = 86.81$~days with the ACF, as indicated by the vertical red dashed lines in the lower panels of Fig.~\ref{fig:HD43587LC}. The modulation of $\sim90$~days is due to the accumulation of discontinuities in the flux data, as can be seen in the upper-right panel (red dots), which creates a false tendency. The second approach was to detrend the entire light curve in each discontinuity, thus eliminating the long-term modulation. We then obtained periodicity peaks at $\sim$7, $\sim$10, $\sim$14, and $\sim$21 days with the GLS, and a peak at 16.96 days with the ACF. However, these GLS peaks do not have any statistical significance as they lay far below the 5\% false alarm probability, as presented in the middle-left panel of Figure \ref{fig:HD43587LC}. 

In accordance with \cite{DoNascimento2014}, such stars should present a rotation period around the solar value, ranging between 14 and 30 days. Even though the rotation period inferred by the ACF method belongs to this interval, the peak corresponding in the ACF curve is not strong enough, compared to the rest, to be considered as a reliable result. Thus, even after eliminating the false long-term tendency, it was impossible to determine a valid modulation period for HD43587 from the CoRoT light curve. \\

We also derived the photometric activity proxy, by computing the standard deviation of the light curve corrected after subtracting the photon noise $S_{ph}$ \citep{mathur2014}. We obtained $S_{ph} = 102$ ppm, which returns a rotational period of  $43.38$ d. This value of $P_{\mathrm{rot}}$ is characteristic of subgiant and RGB stars for this stellar mass, according to \cite{mathur2014}. On the other hand, the weak variability of this light curve, represented by $S_{ph} = 102$ ppm over a 50-year time span, reinforces the flat-activity profile of this star. However, the lack of modulation could also be an indication of a dynamo shutdown. Therefore, these preliminary photometric results strengthen the argument in favor of HD43587 being under a long flat-activity phase.

\begin{figure*}[ht!]
\centering
\includegraphics[width=\textwidth]{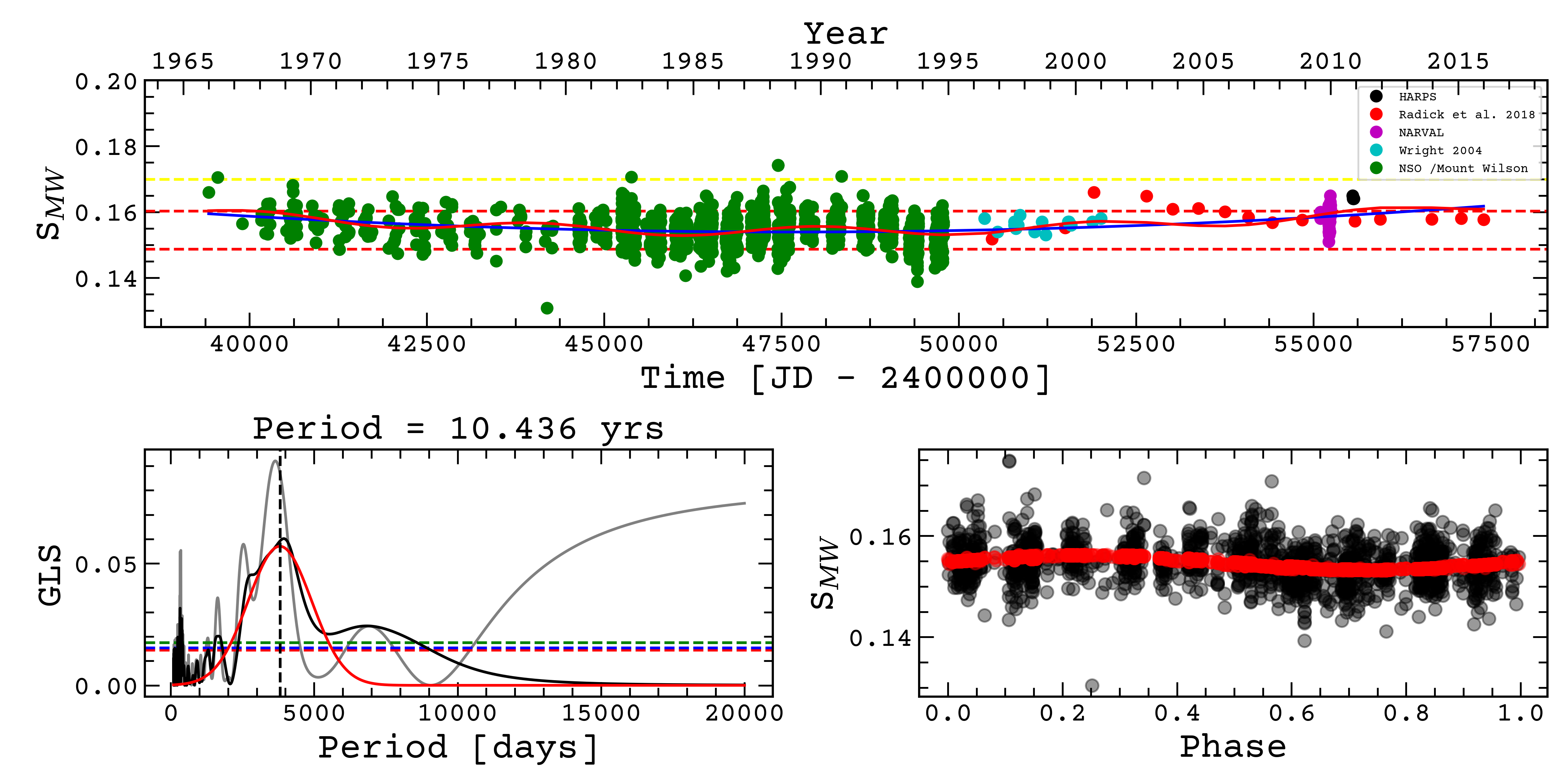}
\caption[All $S$ measurements of HD43587 between 1967 and 2010]
{\emph {Entire spectroscopic chromospheric activity measurements with the $S_{index}$ calibrated to the Mt. Wilson Scale for HD43587 between 1966 and 2016. The upper panel displays all measurements from \citet{duncan}, \citet{wright}, \citet{hall}, and \citet{radick2018}, combined with the computed $S_{MW}$ from the NARVAl and HARPS spectra archives. The red solid line is the sinusoidal curve fitting for 10.436 yrs,  the blue solid line is the long-term trend found to be larger than 50 yrs. The yellow dashed line shows the mean $S_{MW}$ for the Sun. The bottom left panel shows the GLS periodogram of the whole $S_{MW}$ time series (solid gray line), and removing the long trend of over 50 years (solid black line). The Gaussian fit (solid red line) of this second periodogram indicates an activity cycle of 10.436 years (vertical black dashed line). The bottom-right panel shows the phase of the $S_{MW}$ (black circles) and the folded fit with the found period (red circles).}} \label{fig:HD43587Sindex}
\end{figure*}

Finally, we investigated the possibility of revisiting the photometric rotational modulation of this object through a combination of CoRoT and TESS data, in order to enhance the accuracy of the measurements. Unfortunately, after locating this star in the TESS field of view, we realized that HD43587 was situated between two CCDs in one of the TESS field sectors.

\subsection{S-index time-series analysis}
\label{sec:ts}

We computed the S-index for the different spectra obtained with the spectrographs NARVAL and HARPS. We used our own tools developed according to the prescriptions of \citet{wright} and \citet{marsden}. These tools extract the activity index from the Ca II H\&K lines emission fluxes, performing a discrete counting from the spectral lines $H$, $K$, $R$, and $V$. We used Eq.~(\ref{eq.2}) to derive the $S_{MW}$ directly from NARVAL spectra. From HARPS spectra, we computed the instrumental $S_{HARPS}$ following Eq.~(\ref{eq.3}) and calibrated it to the MW scale $S_{MW}$ using Eq.~(\ref{eq.4}).

Afterward, we combined the whole computed $S_{MW}$ time series with the previous published data as described in Sect. \ref{data}. The complete dataset contains 1524 measurements obtained over 50 years between 1966 and 2016. This approach allows us to build a long activity time series calibrated to the MW scale, which is presented in the upper panel of Fig.~\ref{fig:HD43587Sindex}. The standing flat-activity profile, $\braket{S_{MW}} = 0.154$ and $\log \braket{R'_{\mathrm{HK}}} = -4.97$ (see Sect. \ref{sec.sr}), is reinforced by the very low variability of the S-index, quantified by the standard deviation $\sigma_{S}$ = 0.0043.

From this complete activity time series, we investigate the periodicity of a reliable activity cycle of HD43587 consistent with a solar analog star by using the GLS and the Bayesian generalized Lomb-Scargle periodogram with a trend (BGLST) as proposed by \cite{Olspert2018}. The results obtained with both methods are similar and are detailed below.

\subsubsection{Generalized Lomb-Scargle analysis}
\label{sec:gls}

\begin{figure}[ht!]
\centering
\includegraphics[width=\columnwidth]{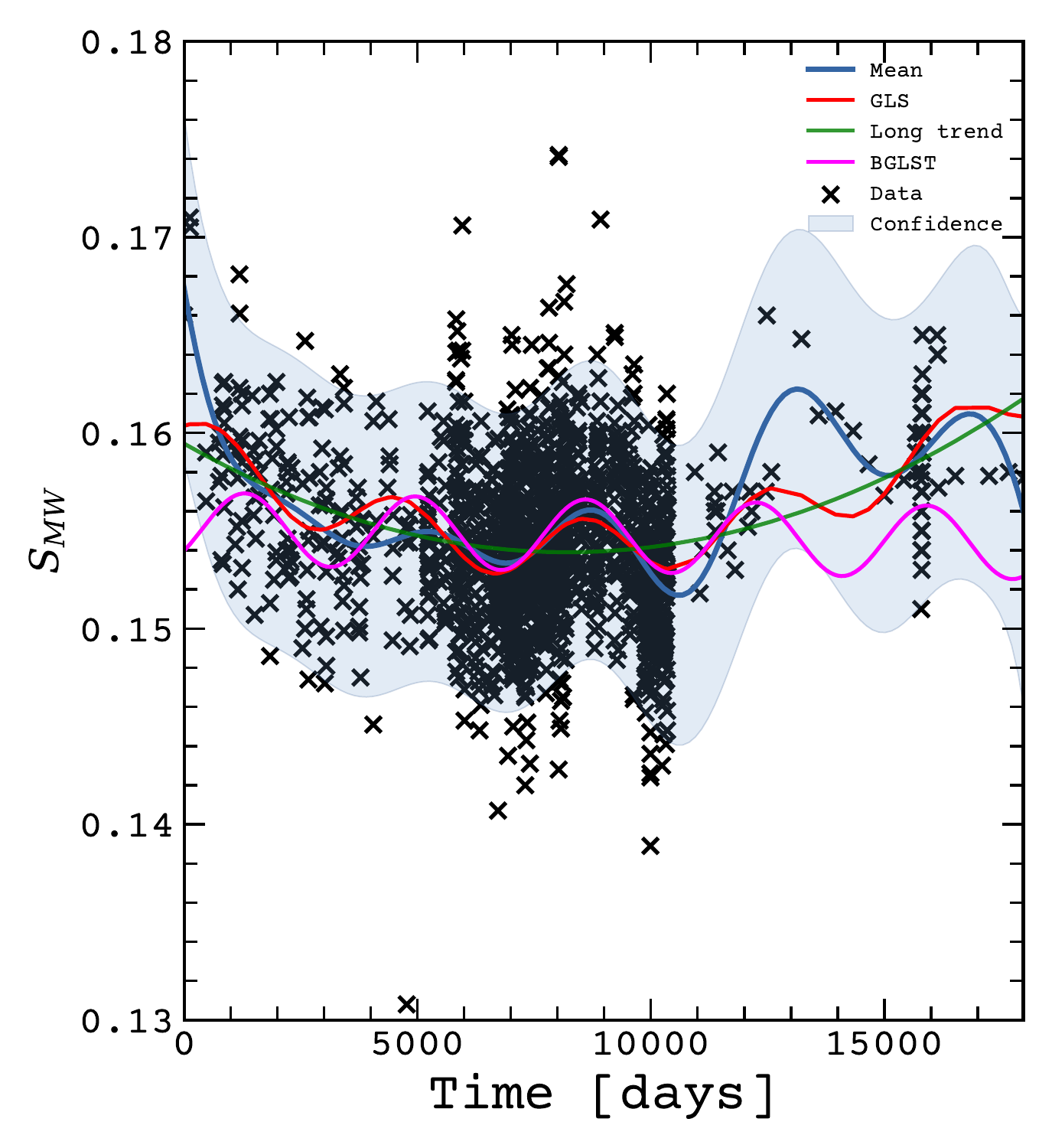}
\caption[$BGLST$ code analysis]
{Comparison between the analysis made with the GLS, GP, and BGLST methods. The green line shows the long trend fit, the red line is a sinusoidal representative fit created with GLS, the magenta  line is the BGLST fit. The blue line represents the GP mean, and the blue shaded region is the confidence region from the GP fit.}
\label{fig:BGLST}
\end{figure}

We used the GLS analysis to compute the activity cycle period, as well as to construct a phase curve for the best determination. The GLS periodogram of the whole $S_{MW}$ time series is represented by the solid gray line in the bottom left panel of Fig.~\ref{fig:HD43587Sindex}. We found two peaks, one around 10 years, and another one, representing a long-term tendency of over 50 years. This latter trend is represented in the upper panel by the solid blue line. We removed this tendency from the time series and generated a new GLS periodogram represented by the solid black line in the bottom-left panel. We used a Gaussian fit (solid red line) to find the cycle period without the long term trend. We thus inferred a cycle period,  $P_{\mathrm{cyc}} = 10.44 \pm 3.03$ yrs. This result is represented by a sinusoidal fitting in the upper panel (red solid line) along with the blue long term trend. The bottom-right panel shows the phase of the folded time series with the sinusoidal model plotted over it. This period of the activity cycle is in good agreement with other solar analog stars \citep[see e.g.,][]{Metcalfe2016}. This result revises the activity cycle length proposed by \cite{egeland_PHD} who, using a shorter dataset, found $P_{\mathrm{cyc}}$ $\approx$ 36 yrs.

We also found a long-term trend that might be a longer cycle with a length exceeding 50 years. This possible longer period could be similar in nature to the period of the solar Gleissberg cycle, which is about 88 yrs. We note that HD43587 could be around the minimum of this cycle as indicated by the blue line in Fig.~\ref{fig:HD43587Sindex}. A considerable observational difficulty lies in the fact that, for HD43587, but also for other analog stars, this secondary cycle may be too long to be determined by using the currently available data. Nevertheless, in the case of HD43587, this result is consistent with the possibility of a primary and a secondary activity cycle \citep{Brandenburg2017}. 

\subsubsection{Bayesian generalized Lomb-Scargle periodogram with trend (BGLST) analysis}

We also performed a periodogram analysis based on the Bayesian generalized Lomb-Scargle method with
linear trend (BGLST), as presented in 
\cite{Olspert2018}. 
When applied to our data, we found a similar period of $\sim$10 years. 
The BGLST method found a linear trend from the beginning of the data until the middle and tried to adjust 
the same tendency until the final fit.  
This “linear fit” moves the final fit a bit upwards. 
Thus, the BGLST method returns a result similar to the GLS analysis presented in the previous section. 
However, instead of removing a long cyclical trend, 
it results in a periodic signal on top of a linear trend 
as shown by the magenta line in Fig.~\ref{fig:BGLST}. 
In addition, we performed  a Gaussian processes (GP) regression with quasi-periodic cycles
by using  the 
GPy code \citep{gpy2014}. 
We did not find any 
meaningful discrepancies on the overall period found by GLS alone. 
Figure \ref{fig:BGLST} shows the 
analysis made with GLS (red line), GP (blue line and shaded blue region), 
and BGLST (magenta line).
The green solid line corresponds to the long trend fit.\\

\subsection{Stellar Rotation}
\label{sec.sr}

Since the photometric analysis using the CoRoT light curve does not
provide a reliable and consistent value for the rotational period of
HD43587, in this section we explore three different alternatives
to compute this parameter. First, we directly used the long-term spectroscopic measurements of $S_{MW}$ over 50 yrs to look for a rotation period in the GLS periodogram. We analyzed the time series from 46340.99 to 46525.65 days (JD-2400000), containing 195 data points spread over 184 days, which represents the denser region of data in the time series. The period values found in the GLS were below the $10 \%$ false alarm probability, and thus discarded as true rotation period.

\begin{eqnarray}
    \centering
    P_{\mathrm{rot}} &=& f(B-V) \cdot g(t)
    \label{gyro1}
\end{eqnarray}

Alternatively, we estimated the rotation period using gyrochronology as proposed by \citet{barnes} and \citet{meibom}, presented in Eq.~(\ref{gyro1}). It consists of computing both the age-dependence function, $g(t) = t^{n}$ (with the age $t$ in Myr) and the mass-dependence function, $f(B-V) = a(B-V-c)^{b}$. The constant terms are $a = 0.770 \pm 0.014$, $b = 0.553 \pm 0.052$, $c = 0.472 \pm 0.027$ and $n = 0.519 \pm 0.007$. We used the two ages determined by \cite{castro}, inferred from two different evolution codes, as presented in Table \ref{table:1} and described in detail in Sect. \ref{HD43587}. The color index used is $B-V$ = 0.61 \citep{VanLeeuwen2007}. Therefore, we found two rotation periods: $P^{T}_{\mathrm{rot}} = 25.0 \pm 4.4$ d with TGEC age, and $P^{C}_{\mathrm{rot}} = 22.9 \pm 3.7$ d with CESTAM age. From these results, we can deduce a theoretical Rossby number, Ro, which is an important parameter of hydromagnetic dynamo theory. It is defined as the ratio of the rotation period, $P_{\mathrm{rot}}$, to the convective turnover time, $\tau_c$, which, in turn, is computed as the ratio of the mixing length to the turbulent velocity one pressure height scale above the base of the convection zone \citep{gilman80}. From the inferred rotation period with the TGEC model's age and convective turnover time $\tau_c = 12.8\pm1.0$ d, we deduce a Rossby number Ro$^{T}= 1.96\pm0.38$. Unfortunately, we had no access to the convective turnover time in CESTAM models and could not infer a Rossby number from them.

In addition, we inferred a rotation period by using the semiempirical calibrations from \cite{noyes84a}, using the time-average observational $S_{MW}$ detailed in the Sect. \ref{data}. We derived the chromospheric contribution, $R'_{\mathrm{HK}}$, from the $S_{MW}$ measurements. It is computed as $R'_{\mathrm{HK}} = R_{\mathrm{HK}} - R_{phot}$, where $\log R_{phot} = -4.02 - 1.40 (B-V)$ and $R_{\mathrm{HK}} = 1.340C_{cf}\cdot S_{MW}$, where $C_{cf}$ is a color-dependent factor proposed by \cite{middelkoop}. \cite{noyes84a} found a cubic fit relation between the Rossby number and the activity index, $R'_{\mathrm{HK}}$ \citep[see Eq.~3 in][]{noyes84a}. We applied this calculation for each value of $S_{MW}$ in our dataset, and determined Ro for each of them. The mean Rossby number with its standard deviation is Ro$^{N}= 2.05 \pm 0.05$. We then determined the convective turnover time by using the empirical relation presented in Eq. (4) of \cite{noyes84a}, which depends on $(B-V)$. With the resulting value, $\tau_c = 9.67$ d, and the definition of the Rossby number Ro = $P_{\mathrm{rot}}/\tau_c$, we derived a mean $P^{N}_{\mathrm{rot}} = 19.8 \pm 0.7$ d. 
 
With these results, we infer an average rotation period, $\overline{P}_{\mathrm{rot}} = 22.6 \pm 1.9$ d for HD43587, which is consistent with the rotation of a solar analog. The good estimate of $\overline{P}_{\mathrm{rot}}$ found is reinforced when we examine the ratio of the cycle frequency to the rotation frequency given by $\omega_{\mathrm{cyc}} / \Omega = P_{\mathrm{rot}} / P_{\mathrm{cyc}}$ as suggested by \citet{Brandenburg2017}. In the case of the primary cycle, $P_{\mathrm{cyc}} = 10.44$ yrs, we have $\log (\overline{P}_{\mathrm{rot}} / P_{\mathrm{cyc}}) = -2.23$, while the solar value is $\log (P_{\mathrm{rot}} / P_{\mathrm{cyc}})_{\odot} = -2.20$. In the case of the possible secondary cycle, $P_{\mathrm{cyc}} \geq	50$ yrs, we found $\log (\overline{P}_{\mathrm{rot}} / P_{\mathrm{cyc}}) \leqslant -2.87$. For the Sun, the long Gleissberg cycle (GC) gives a ratio $\log (P_{\mathrm{rot}} / P_{\mathrm{cyc}})_{\odot,GC} = -3.10$. If we examine the position of HD43587 in Fig. 5 of \citet{Brandenburg2017}, which plots this ratio as a function of $\log \braket{R'_{\mathrm{HK}}}$, we can see that 
HD43587 lies on the inactive branch region, highlighting that the $\log \braket{R'_{\mathrm{HK}}} = -4.97$ value is below the Vaughan-Preston gap ($\log \braket{R'_{\mathrm{HK}}} = -4.75$).

Finally, another interesting fact concerning HD43587 is that 
its Rossby number,
as determined above, is close to the critical Rossby number, Ro $\sim 2$, beyond which the efficiency of magnetic braking is dramatically reduced \citep{VanSaders2016,Metcalfe2016}. This region corresponds to an activity level 
$\log R'_{\mathrm{HK}} \sim -4.95$. 
This suggests that HD43587 lies in a zone below the Vaughan-Preston gap where the differential rotation, key mechanism of the dynamo process, becomes inefficient 
\citep{Metcalfe2016}.

\begin{figure}[t!]
    \centering
    \includegraphics[width=\linewidth]{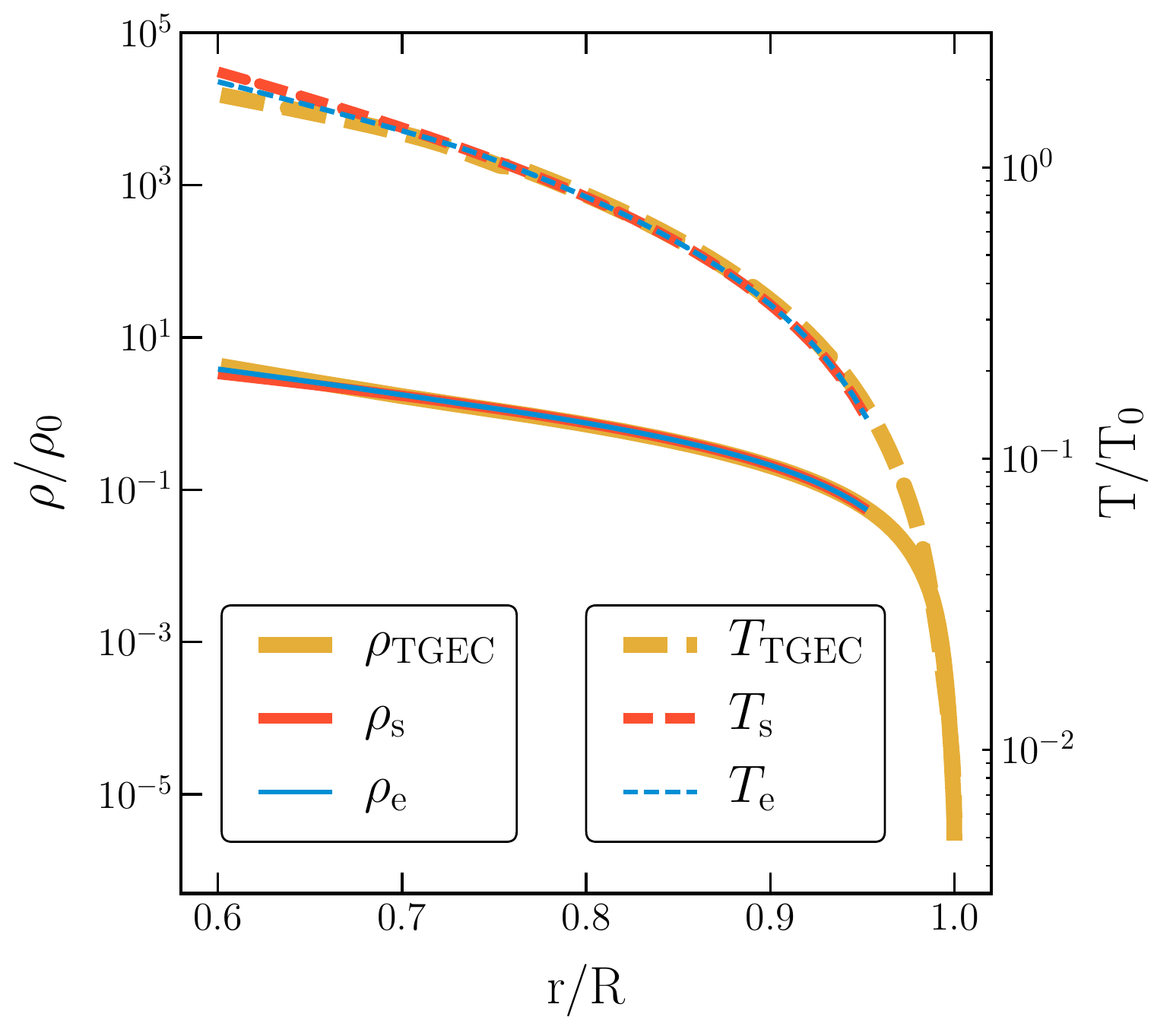}
    \caption{Radial profiles of the temperature, $T$ (dashed lines), 
    and density, $\rho$ (solid lines). The blue and red lines correspond 
    to the ambient and isentropic states, respectively. The yellow lines 
    show the TGEC stratification.}
    \label{fig:strat}
\end{figure}

\section{Simulating HD43587 with the EULAG-MHD code}

A relevant tool currently available for studying stellar magnetism is 
global numerical modeling. Several recent works have considered this approach 
to study  the magnetic 
activity of solar-like stars rotating at different rates
\citep[e.g.,][]{BMBBT11,ABT14,SBCBdN17,warnecke18,Viviani+18,G+19a}. The results
have provided a new understanding of the relation between stellar rotation and 
magnetic activity. To substantiate our observational analysis we also perform numerical simulations of an atmosphere that mimics the stratification of HD43587. Our goal is not to make a detailed analysis of the simulations, but to get some insights about the internal activity of this star providing support to the observational conclusions.

\subsection{Description of the models}

We simulated the internal dynamics of the star HD43587 with the EULAG-MHD 
code \citep{SC13}.  The fundamental difference between the models presented 
here and previous solar global dynamo simulations is the adoption of
a fiducial fit for its stratification. Thus, structure parameters as the 
location of the tachocline or the velocity field in the convection zone
are properly captured.  The code solves the anelastic 
equations of \cite{LH} extended
to the MHD case and in the inviscid regime. 
One relevant point distinguishing EULAG-MHD simulations from global
dynamo modeling with other codes is the use of the implicit subgrid
scale (SGS) method MPDATA. It allows us to keep the numerical viscosity sufficiently
low in such a way that it is possible to resolve the transition between convective 
and radiative zones without much transfer of angular momentum. Within the 
convection zone, the numerical dissipation introduced by the MPDATA
scheme plays the role of the SGS contribution to the large scale
motions and magnetic field.

We consider a spherical shell, $0 \le \varphi \le 2 \pi$, $0 \le \theta \le \pi$, 
radially spanning from $r=0.6R$ to $r=0.96R$, and a mesh 
resolution of $128\times 64 \times 48$ grid points in $\varphi$, 
$\theta$ and $r$, respectively. 
The full set
of equations are described in \cite{G+16}. The boundary conditions for
the velocity field are impermeable and stress free at both boundaries. 
For the potential temperature perturbations we consider zero radial 
derivative of the radial convective flux at the bottom, and zero convective radial flux at the top. The magnetic boundary conditions correspond
to a perfect conductor at the bottom and pseudo-vacuum at the top.  The
simulations start with white noise perturbations in the potential
temperature, the radial velocity, and the azimuthal magnetic field.   
Since the main input parameter of the model is the stellar stratification, we present details of its setup below. 

As described in \cite{G+16} and \cite{C+17}, two states are necessary in
the formulation of convective simulations with EULAG-MHD: the isentropic reference state, 
and the ambient state. The later allows adiabatic displacements of fluid parcels in the super-adiabatic layers, and inhibits convective motions where the ambient is
sub-adiabatic. Both states are described by hydrostatic equilibrium
equations:

\begin{eqnarray}
    \centering
    \frac{\partial T_{\rm i}}{\partial r} &=& -\frac{g}{R_g (m_{\rm i}+1)}\;,
    \label{equ:equilT}
    \\[10pt]
    \frac{\partial \rho_{\rm i}}{\partial r} &=& -\frac{\rho_{\rm i}}{T_{\rm i}}
     \left(\frac{g}{R_g} - \frac{\partial T_{\rm i}}{\partial r} \right)\;,
     \label{equ:equilrho}
\end{eqnarray}

\noindent
where the index $i$ stands either by $s$, for the reference state, or by $e$, for the 
ambient state. In the reference state, the polytropic index $m_s$ is equal to 1.5.
The ambient stratification is a bi-polytropic fit to the mixing length (ML)
stratification model obtained with the evolutionary code TGEC as described by \cite{castro}. 

\begin{figure}[t!]
    \centering
    \includegraphics[width=\linewidth]{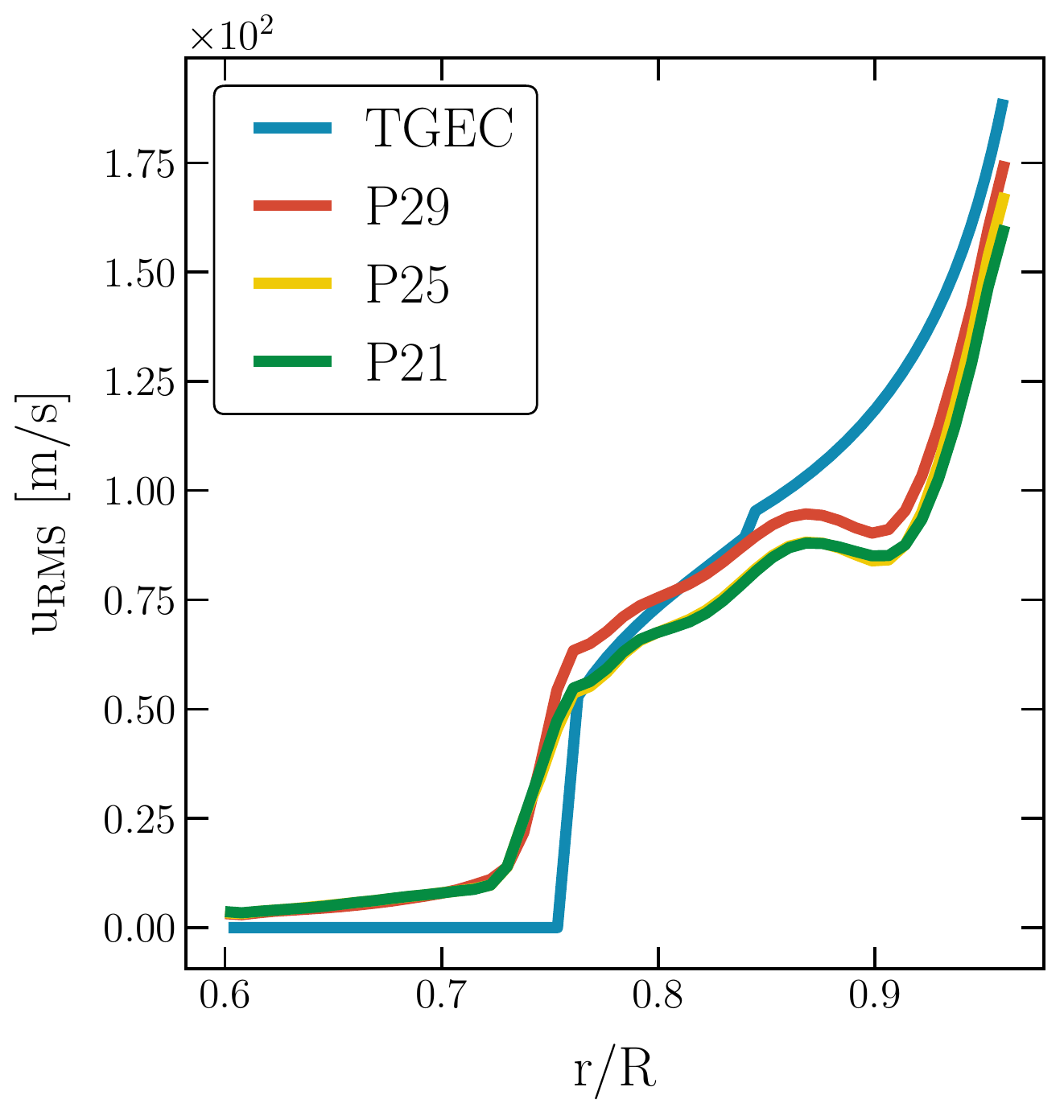}
    \caption{Radial profile of the turbulent velocity for simulations P29 (red), P25 (yellow) and P21 (green). The blue line shows the TGEC stratification model.}
    \label{fig:turbvel}
\end{figure}

Our fit takes into account the recent results of \cite{G+19a}, indicating that the interface between the radiative and convective zones might play a relevant role in the dynamo generating mechanism. Their results also show that the magnetic field is stored in the stable layer where it is prone to magnetoshear instabilities \citep[see e.g.,][]{Miesch+07}, in turn developing helical non-axisymmetric motions and currents (source of magnetic field via the so-called $\alpha$-effect). As demonstrated in \cite{G+19b}, 
these instabilities are highly dependent on the stratification of the atmosphere. 
For this reason we calibrate the ambient state to reproduce the Brunt–Väisälä
frequency profile of TGEC in the stable stratified layer. Since the motions
in the convection zone define the development of the differential
rotation, in this region we adjust the polytropic index allowing
for a radial profile of the turbulent velocity compatible with that of
the TGEC model (see Fig.~\ref{fig:turbvel}). Thus,  the quantity $m_e$ in 
Eq.~(\ref{equ:equilT}) is given by

\begin{eqnarray}
    \centering
    m_e(r) = m_{rz} + \frac{\Delta m}{2} \left[1 + {\rm erf} \left( \frac{r - r_{tac}}{w_t} \right) \right]
\end{eqnarray}

where $m_{rz} = 2.0$ is the polytropic index of the radiative zone, $\Delta m = m_{cz} - m_{rz}$, 
with $m_{cz} = 1.49989$ the polytropic index in the convective zone, and $w_t = 0.1 R$ the width of transition 
between both regions. The radiative zone extends until $r_{tac}=0.76R$, the convective zone corresponds 
to the upper envelope. The profiles of temperature and density are 
shown in Fig.~\ref{fig:strat}. The yellow solid and dashed lines depict the ML stratification from TGEC model, and the blue and red solid (dashed) lines 
correspond to the fitted density (temperature) profile for the ambient 
and reference states, respectively.

We performed three simulations with the same stratification and different 
rotational 
period. They are labeled as P29, P25, and P21 for 29, 25, and 21 days,
respectively. 
The rotation periods cover the uncertainties in the determination of $P_{\mathrm{rot}}$ from the analysis presented in \S \ref{sec.sr}. In Fig.~\ref{fig:turbvel}, we compare the 
convective velocity from TGEC model with the turbulent velocity, $u_{\rm RMS}$ obtained  in the
three global simulations performed with EULAG-MHD. 
Since rotation constrains the convective motions, 
the amplitude of $u_{\rm RMS}$ 
decreases with the decrease of $P_{\mathrm{rot}}$.  All three 
velocity profiles exhibit, however,  the same trend. The values of $u_{\rm RMS}$ volume averaged in
the convection zone are presented in Table~\ref{tab:2}.
The Rossby number presented in the same table was computed as it was in
\cite{gilman80}, with $\tau_c$ computed from the pressure scale
height and the resulting  $u_{\rm RMS}$ one pressure scale height
above the star's tachocline.  At this radial point, the turbulent
velocity in simulations P25 and P21 is smaller than the one predicted by 
the mixing length model.
Thus, Ro is smaller than the one obtained from the TGEC model for the same rotational period, but still within the determination error of Ro$^T$. 

Also in contrast with TGEC, the turbulent velocity in the simulations is different from zero in the radiative zone. These motions have two origins, overshooting of the downward flows coming from the convection zone, and non-axisymmetric motions developed in the stable layer due to MHD instabilities. As we present below, the realistic stratification of HD43587 captured in the ambient state  allows us to reproduce some aspects of the magnetism of this star which are compatible with the  observational results obtained in the previous sections.

\subsection{Simulation results}

\begin{figure}
    \centering
    \includegraphics[width=\linewidth]{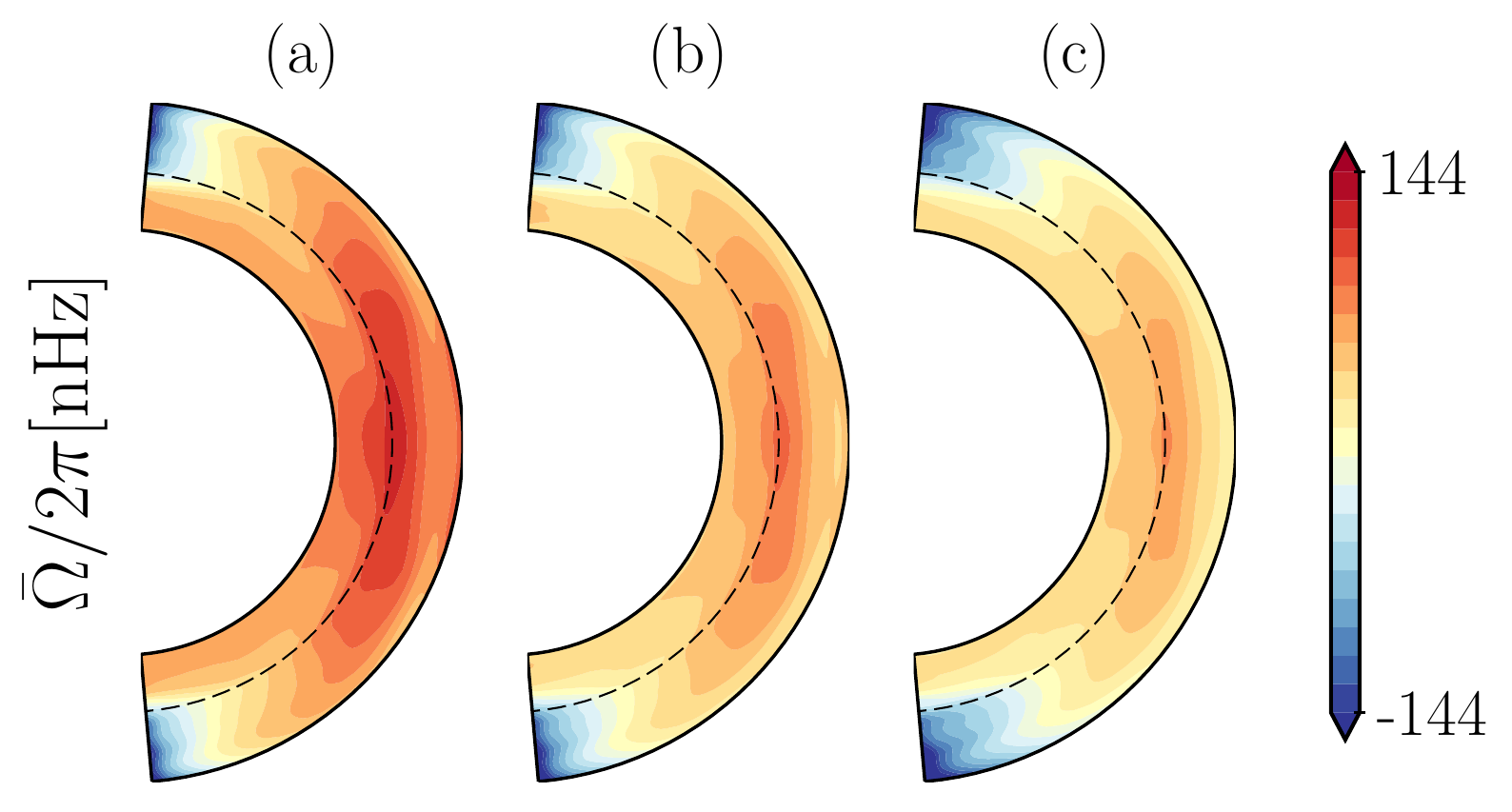}
    \caption{Mean differential rotation in the meridional plane 
    from simulations (a) P29, (b) P25 
    and (c) P21. The color scale shows the angular velocity 
    with respect to the 
    rotation frame. The dashed lines depict the tachocline location.}
    \label{fig:omega}
\end{figure}

Figure~\ref{fig:omega} shows the differential rotation profiles resulting from 
the simulations (a) P29  to (c) P21.
The mean angular velocity, $\bar{\Omega}$, is an average on longitude and 
time (along the last two magnetic cycles of each simulation).
All the cases show a clear formation of a radial shear layer at 
the base of the convection zone. 
For the three rotational periods studied, the resulting differential rotation is 
solar-like, with slow poles and a fast equator. 
This kind of differential rotation appears when the timescale of 
the convective flow is similar to the rotational one, meaning
rotationally constrained motions. In this case, the transport of
the angular momentum is directed towards the equator \citep[see e.g.,][]{FM2015,Guerrero20}.

In the simulations P29-P21, we notice that the latitudinal gradient 
of angular velocity diminishes as the rotational period decreases 
(Fig.~\ref{fig:omega} (a) to (c)) This is quantitatively verified in the third 
column of Table~\ref{tab:2} showing the relative differential rotation, $\Delta \Omega/\Omega_{eq}$, 
where $\Delta \Omega = \Omega_{eq} - \Omega_{45}$. In all the cases, this
value is smaller than the solar one, $\sim 0.1$, and about one order
of magnitude smaller than in other stars where it was inferred by 
astroseismology \citep{Benomar+18}.

In Fig.~\ref{fig:butterfly}, the colored contours depict the radial magnetic field
averaged over longitude, $\overline{B}_r$, and the solid (dashed) contour lines represent
the positive (negative) mean toroidal magnetic field, $\overline{B}_{\varphi}$. 
In general, the simulations result in oscillatory magnetic fields with polarity 
reversals.  Like in \cite{G+19a}, the results suggest the existence of
a $\alpha^2-\Omega$ dynamo operating inside HD43587.
The magnetic fields are anti-symmetric across the equator. For slow
rotation the field seems further and further concentrated near the poles.
The averages in volume and time of the toroidal and radial
fields are also presented in Table~\ref{tab:2}. We chose to present
$\langle \overline{B}_r \rangle$ instead of the poloidal field because 
the radial field is directly observed in the Sun.
\begin{figure}[t!]
    \centering
    \includegraphics[width=\linewidth]{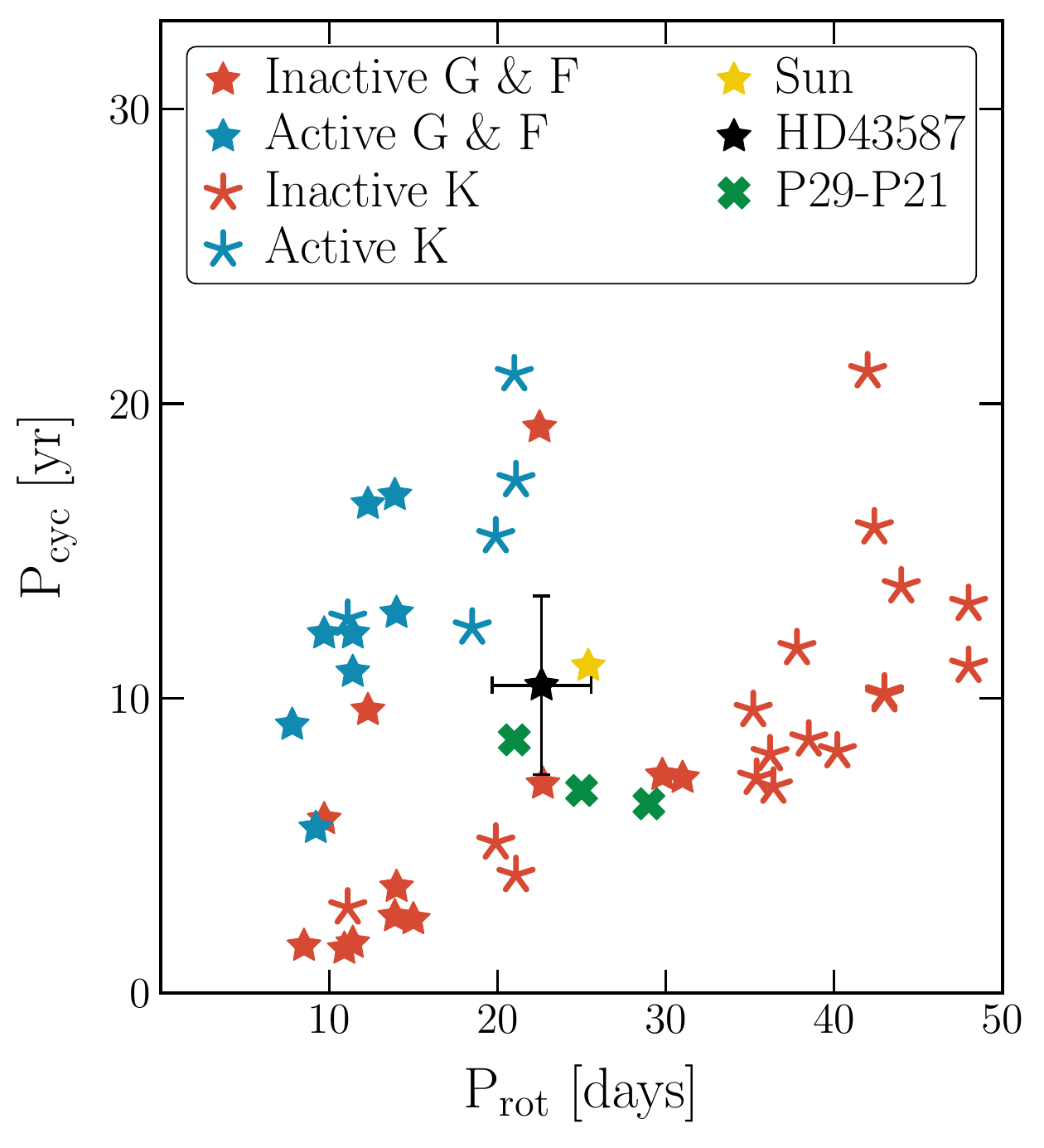}
    \caption{Correlation between $P_{\rm cyc}$ and $P_{\rm rot}$ for the
    simulations presented in Table~\ref{tab:2} (green points). The blue and red points
    correspond to F, G, and K star data as presented in \cite{Brandenburg2017}. The black star shows the position of 
    HD 43587 with the values inferred in this work, and the yellow point marks the position of the Sun. }
    \label{fig:prpc}
\end{figure}

The magnetic cycle periods, $P_{\rm cyc}$,  were computed by Fourier analysis (see Table~\ref{tab:2}).
The values, between 6.4 and 8.6 years are slightly smaller than that of 10.4 years obtained in the chromospheric activity analysis . Nevertheless, the cases for 
21 and 25 days are 
still within the $1\sigma$ error of 3.3 years. In Fig.~\ref{fig:prpc} we show the results of the simulations in the $P_{\rm rot}$ versus $P_{\rm cyc}$ plane.
The figure contains the observational points of F, G, and K stars as presented in 
\cite{Brandenburg2017}. The rotational and magnetic cycles of HD43587 are
presented with a black star. The periods of the three simulations fall in the 
inactive branch of activity closer to the position of the Sun. The agreement
between observation and simulation is encouraging, and
indicates that the simulations are capturing the physics of the interior of 
the star.

In the recent work of \cite{Olspert2018}, the results of novel
analysis of the MW data suggest that the active branch presented in 
Fig.~\ref{fig:prpc} may not exist. Instead, they found a branch merging 
the active and transitional branches. Although this might be the case,
we consider that there are not yet sufficient stars observed with fiduciary
determination of their periods. Thus, while the activity-rotation problem
remains a subject of discussion \citep{BoroSaikia2018}, we decided to present our results
together with the canonical A and I branches.

\begin{figure*}[t!]
    \centering
    \includegraphics[width=\textwidth]{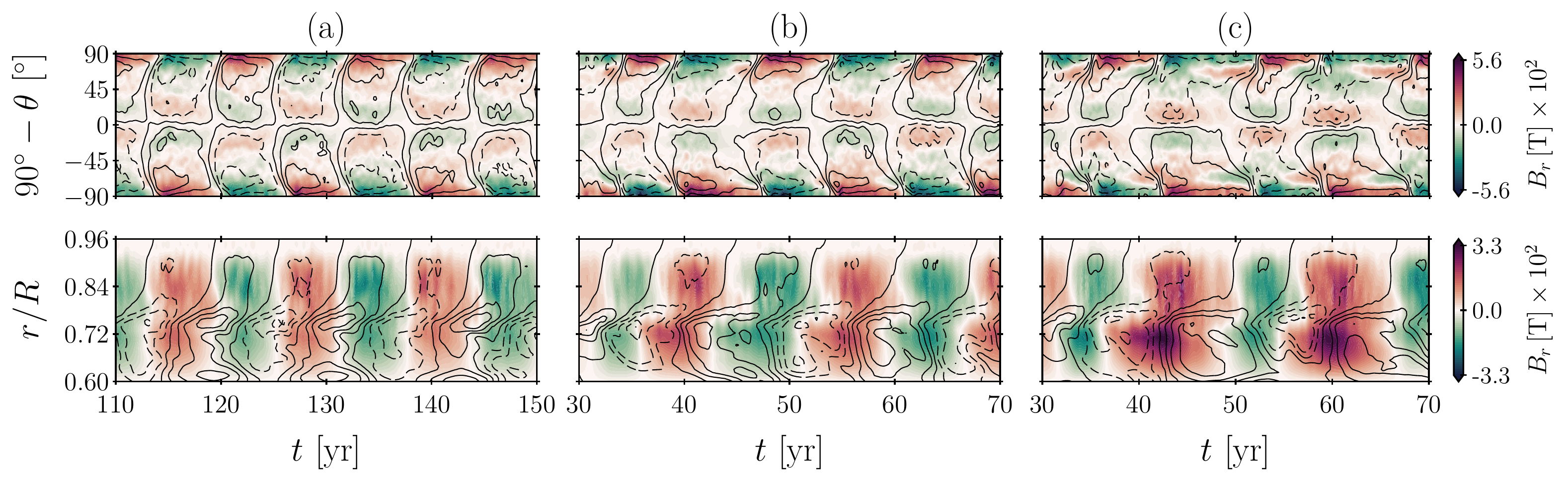}
    \caption{Time latitude at $r=0.90\, R$ (upper panels), and time radius
    at  $\theta=15^{\circ}$ latitude (lower panels), butterfly diagrams comparing the magnetic field evolution between simulations (a) P29, (b) P25, and (c) P21.  The colored contours show the radial magnetic field intensity. The solid (dashed) lines show the positive (negative) contours of toroidal magnetic field.}
    \label{fig:butterfly}
\end{figure*}

The amplitude of the toroidal magnetic field in 
the generating radial shear layer is about $0.4$ Tesla and is larger 
in the simulations with a larger period. 
The radial field component, on the other hand, is about $0.01$ Tesla and is
larger in the simulations with a shorter period (fast rotating).  Figure~\ref{fig:butterfly} indicates
that the magnetic field migrates upwards, however, at the upper
boundary of the model, $r=0.95R_{\star}$, both, the toroidal and the
radial field components decay about one order
of magnitude. Furthermore, the simulation results do not exhibit relevant 
dynamo action near the upper boundary. Thus, it is expected that it will 
decrease further when reaching the star surface, which is not considered in 
the model.

In summary, the simulations are able to reproduce cyclic dynamo action,
including field reversals, with a differential rotation smaller than in 
the Sun, and a magnetic field, which at the top boundary is similar to
the radial field observed at the solar surface ($\sim 10^{-3}$ Tesla). 
The 10\% difference observed in  
$\langle S_{MW} \rangle$  between the Sun and HD43587 might be correlated
to a diminished differential rotation, which in the Sun is at least twice
as large as in simulations P25 and P21. These cases are the ones that better reproduce  the cycle period of HD43587 inferred from the spectroscopic 
chromospheric activity.


\section{Conclusions and summary}

The spectral analysis is fundamental to understanding 
astrophysical processes 
and stellar evolution. Currently, the stellar magnetic activity is one of the most important observables in the context of stellar interiors, and it
might
impact exoplanet detection as shown by \citet{aigrain}. In this work, from the high-resolution spectra archive and already published data, we 
investigated the chromospheric activity evolution of the solar analog star HD43587. 

\begin{table*}[h]
    \centering 
    \caption{Simulation parameters and results. }
    \label{tab:2}
    \begin{threeparttable}
    \begin{tabular}{c c c c c c c c c c c}
        \hline\hline
        \\
        Model & $P_{\mathrm{rot}}$ & ${\rm min}(Ra)$\tnote{1}  & ${\Delta \Omega}/{\Omega_{eq}}$\tnote{2} & $\langle u_{RMS} \rangle$\tnote{3} &  $Ro$\tnote{4} & $P_{cyc}$ & $\langle \overline{B}_{\phi} \rangle_{NSL}$\tnote{5} & $\langle \overline{B}_{r} \rangle_{NSL}$\tnote{5} & $\langle \overline{B}_{\phi} \rangle_{TAC}$\tnote{5} & $\langle \overline{B}_{r} \rangle_{TAC}$\tnote{5} \\
          & [days] &  & [m s$^{-1}$] &  & [years] & [T] & [T] & [T] & [T] \\
        \hline
        \\
        P29 & 29 & $4.5\cdot10^6$ & 0.087 & 87.06 & 2.13 & 6.42 & 0.0422 & 0.0012 & 0.4559 & 0.0113 \\
        P25 & 25 & $4.5\cdot10^6$  & 0.047 & 80.13 & 1.64 & 6.87 & 0.0403 & 0.0013 & 0.4372 & 0.0147 \\
        P21 & 21 & $4.5\cdot10^6$  & 0.023  & 79.35 & 1.38 & 8.59 & 0.0392 & 0.0016 & 0.4027 & 0.0174 \\
        \hline
    \end{tabular}
        \begin{tablenotes}
        \item[1] A minimum Rayleigh number for the simulations may be estimated as ${\rm min}(Ra) = \frac{g_0 r_{\rm cz}^2}{\nu \alpha \Theta_e} \frac{d \Theta_e}{d r}$, where $\Theta_e$ is the ambient state potential temperature,         $\alpha=9.64\times10^{-9}$ s$^{-1}$, is the inverse of the Newtonian cooling time limiting the growing of perturbations of $\Theta$, $r_{\rm cz} = 0.2R_{\star}$ is the depth of the convection zone and $\nu = 10^8$ m$^2$ s$^{-1}$ is the maximum value of the numerical viscosity estimated by \cite{SBBCMS16}, for EULAG-MHD simulations with a similar resolution.
        \item[2] Differential rotation parameter, $\Delta \Omega /\Omega_{eq} = 
        (\Omega_{eq} - \Omega_{45})/\Omega_{eq}$, where  $\Omega_{eq}$ and $\Omega_{45}$ are the angular velocities at $0^{\circ}$ and  $45^{\circ}$ latitude, respectively. 
        \item[3] $\langle u_{RMS} \rangle$ is the volume average of the turbulent velocity in the entire convection zone. 
        \item[4] Rossby number is defined as ${\rm Ro}=P_{\mathrm{rot}}/\tau_c$ and computed with $\tau_c = \alpha H_p / u_{RMS}$, where $H_p$ is
        the pressure scale height, and it is measured at one pressure height scale above the bottom of the convective zone \citep{gilman80}. 
        \item[5] Toroidal and radial magnetic field averaged over longitude and time. The subscripts $NSL$ and $TAC$ correspond to $r=0.95R_{\star}$ and
        $r=0.76R_{\star}$, respectively.
        \end{tablenotes}
    \end{threeparttable}
\end{table*}{}

We built the larger $S_{MW}$ time series already published for this star, using data spanning 50 years from 1966 until 2016. The derived $\braket{S_{MW}} = 0.154$ and $\log \braket{R'_{\mathrm{HK}}} = -4.97$ 
indicate 
a low chromospheric activity, beyond the Vaughan-Preston gap. From the GLS analysis of the $S_{MW}$ time series, we found a reliable 10.4 yr cycle period (see Fig. \ref{fig:HD43587Sindex}), 
which is similar
to the solar value. The indirect determinations of a rotation period give $\overline{P}_{\mathrm{rot}} = 22.6 \pm 1.9$ d. 
These values of
$P_{\rm cyc}$ and $\overline{P}_{\mathrm{rot}}$,
are consistent for a solar analog star. The inferred Rossby number, Ro$^{N}$ = $2.05\pm0.05$ from the calibrations prescribed by \citet{noyes84a}, and Ro$^{T}$ = $1.96\pm0.38$ from TGEC modeling, point out a rotation period about two times larger than the convective turnover time which
might represent a diminished differential rotation 
\citep{Metcalfe2016}.

Based on the inferred rotational period and the internal structure
of HD43587, we performed global simulations with the EULAG-MHD code.
We considered three different rotational periods around the value 
obtained in \S\ref{sec.sr}. For all the cases, the angular velocity 
is solar-like, i.e., fast equator and slow poles, with relative differential 
rotation smaller than the solar value.  In principle, the smaller shear 
might be responsible for the low levels of activity of the
simulated stars. The 
radial field at the top of the simulations (at $r=0.96\,R_{\star}$) is
about $10^{-2}$ T.  This value is similar to the radial field at the
solar photosphere.

The magnetic field is oscillatory with magnetic field reversals. The
magnetic cycle periods obtained in the simulations  range from
$6.4$ yrs for the simulations with a rotational period of 29 days, to
$8.6$ yrs for the simulation with $P_{\rm rot} = 21 $ days.  This cycle
period is compatible with the one inferred in \S\ref{sec:ts}.

It is worth mentioning that given the difficulties of finding the rotation period from the light curves, the indirect determination of the $P_{\rm rot}$ is the best approach to to characterize the star HD43587. The fact that the inferred rotation period fits with a solar analog encourages us to presume that it might be correct. Furthermore, ab-initio simulations of the star with initial parameters given by the stellar structure (the same as used to determine the rotational period) as well as the inferred $P_{\rm rot}$, result in activity cycles close enough to that independently determined from the $S$-index. These numerical results substantiate the observational ones.

Even though the possibility that HD43587 is under an MM cannot be ruled out, all our results seem to indicate that it is a solar analog star, older than the Sun, and undergoing a "natural" decrease in its magnetic activity. Nevertheless, some authors argue that a transition in the stellar differential rotation might be the cause of these low levels of activity \citep{MS17}. The characterization of an MM would only be valid if we could observe the star coming in and out of this activity minimum \citep{hall}. There is no current observation indicating this pattern. 

The low activity level observed in HD43587 might be due to a small differential rotation in the convective zone. We have also found evidences that HD43587 could be at the minimum of a longer activity cycle of over 50 years. However, in the context of observations of magnetic cycles in solar-like stars, the existence of long activity cycles for old objects is rather controversial. The results presented by \cite{Brandenburg2017} show that, besides the Sun, all confirmed objects that have co-existing long and short cycles are younger than 2.3 Gyr. As a matter of fact, the reality of double periodicities in the observed stellar activity cycles is still questioned within the community \citep{BoroSaikia2018,Olspert2018}.

The results presented above demonstrate the relevance of studying HD43587 in the context
of stellar evolution. Several authors have investigated its features and evolutionary status, as well as its magnetic activity evolution. Its similarity with the Sun, whether evolutive, spectroscopic, or magnetic, gives insight to the future evolution of the Sun and makes it a good candidate for being an Earth-like exoplanet host. Its low activity level provides an excellent opportunity to attempt a signal of an earth-like planet without the interference of the magnetic activity \citep{Korhonen2015}. In this way, we estimate that it is fundamental to keep monitoring the activity of HD43587 to confirm the co-existence of a short and long period cycle, as well as to increase the accuracy of the cycle and rotation period determination. We also consider that it should be a key target for the ESA PLATO mission.

\begin{acknowledgements}
We thank the anonymous referee for the useful comments and suggestions that improved this paper.
Research activities of the Ge$^3$\footnote{\url{http://ge3.dfte.ufrn.br}} stellar team at the Federal University of Rio Grande do Norte are supported by continuous grants from Brazilian scientific promotion agencies CAPES and CNPq. R.R.F. and R. B acknowledge CAPES and CNPq 
graduate fellowships. J.D.N. and M.C. acknowledge support from CNPq ({\em Bolsa de Produtividade}). We also thank T. Morel for providing the HARPS spectra used in this paper. This work made use of the computing facilities of the Laboratory of Astroinformatics (IAG/USP, NAT/Unicsul), whose purchase was made possible by FAPESP (grant 2009/54006-4) and the INCT-A.

\end{acknowledgements}

\bibliographystyle{aa}
\bibliography{bibliography.bib}

\end{document}